# Numerical models for AC loss calculation in large-scale applications of HTS coated conductors

**Loïc Quéval[1], Víctor M. R. Zermeño[2], Francesco Grilli[2]**

[1]University of Applied Sciences Düsseldorf, Josef-Gockeln Strasse 9, 40474 Düsseldorf, Germany
Phone: +49-2114351318, e-mail address: loic.queval@gmail.com

[2]Karlsruhe Institute of Technology, Hermann-von-Helmholtz Platz 1, 76344 Eggenstein-Leopoldshafen, Germany
Phone: +49-72160828528, e-mail address: francesco.grilli@kit.edu
Phone: +49-72160828582, e-mail address: victor.zermeno@kit.edu

**Abstract**

Numerical models are powerful tools to predict the electromagnetic behavior of superconductors. In recent years, a variety of models have been successfully developed to simulate high-temperature-superconducting (HTS) coated conductor tapes. While the models work well for the simulation of individual tapes or relatively small assemblies, their direct applicability to devices involving hundreds or thousands of tapes, as for example coils used in electrical machines, is questionable. Indeed the simulation time and memory requirement can quickly become prohibitive. In this article, we develop and compare two different models for simulating realistic HTS devices composed of a large number of tapes: 1) the homogenized model simulates the coil using an equivalent anisotropic homogeneous bulk with specifically developed current constraints to account for the fact that each turn carries the same current; 2) the multi-scale model parallelizes and reduces the computational problem by simulating only several individual tapes at significant positions of the coil's cross-section using appropriate boundary conditions to account for the field generated by the neighboring turns. Both methods are used to simulate a coil made of 2000 tapes, and compared against the widely used H-formulation finite element model that includes all the tapes. Both approaches allow speeding-up simulations of large number of HTS tapes by 1-3 orders of magnitudes, while keeping a good accuracy of the results. Such models could be used to design and optimize large-scale HTS devices.

## I. Introduction

The high current capacity of high-temperature-superconducting (HTS) conductors makes them ideal candidates for compact and powerful electromagnetic devices such as cables, magnets, motors, generators and SMES. While most of these devices are designed to operate in DC conditions, for many applications, ripple fields are expected as part of the normal operation of the device. Furthermore, several devices such as AC dipole magnets or cables are designed for working in AC conditions. As a result of these ripple and AC fields, hysteretic losses (AC losses) are expected in both transient and steady-state operation. Estimating and understanding these losses is important for performance evaluation and design.

For a given device using HTS coated conductor tapes, this reduces to calculating hysteretic losses of a stack or an array of tapes for at least one period. HTS material is usually modeled with a power law *E-J* relationship that includes the dependence of $J_c$ on the magnetic flux density, hence taking overcritical current densities into consideration. This nonlinear property turns the calculation of hysteretic losses into a challenging task for devices made of hundreds or thousands of tapes since both large computing time and memory are required [Sirois2015].

In this context, numerous models have been proposed to simulate large stacks and arrays of HTS coated conductors {in the references that follow, the maximum reported number of simulated tapes is indicated in brackets}. A first group of models uses the infinitely thin film approximation: T-formulation finite element model [Ichiki2005]{3}; 1D integral equations model [Brambilla2009]{4}. A second group of models considers the exact conductor geometry: quasi-variational inequalities formulation [Souc2009]{32}; H-formulation finite element model with large aspect ratio mapped mesh in the superconductor layer [Zermeno2011]{57}; minimum magnetic energy variation model assuming uniform current distribution for the tapes far from the tape of interest [Pardo2013]{6400}, integral equations model assuming that the tapes far from the tape of interest can be grouped and carry a uniform current distribution [Carlier2012]{12}. A third group of models uses an anisotropic homogeneous-medium approximation: [Clem2007]{n/a}, [Yuan2009]{1000}, and [Zhang2014]{1200} are based on the critical state model and use a predefined distribution of current in the stacks under analysis by means of either a straight line or a parabola to manually separate regions with positive, negative or zero current density. A model that does not make any a priori assumptions about the current distribution in the homogenized stack is shown in [Prigozhin2011]{100}, however, being also based on the critical state method, this work does not consider overcritical currents. A later model that does not make any a priori assumptions about the current distribution and that allows for overcritical local currents was presented in [Zermeno2013b]{64}. A fourth group of models uses a multi-scale approach: first the background field is obtained by assuming that the current density in the coil windings is uniform, then the AC losses in each conductor is estimated. The estimation can either rely on measurements of single conductors [Oomen2003]{2018}, or on numerical calculations [Tonsho2004]{2000}, [Queval2013]{40204}.

In this article, we assess and compare the ability of two different finite element models in simulating arrays made of a large number of tapes. The first model is an extension of the homogenized model introduced in [Zermeno2013b], which simulates the coil using equivalent anisotropic homogeneous bulks with specifically developed current constraints to account for the fact that each turn carries the same current. New results confirm the effectiveness of such model in simulating large arrays of coated-conductors, with reduced efforts. The second model is a multi-scale model inspired by [Queval2013]. The main idea here is to parallelize the problem by breaking up the computational domain into several smaller domains using appropriate boundary conditions. Hence the problem of simulating large coils reduces to simulating only one conductor at a time. This in turn reduces the memory requirement and the overall computational burden. Here the convergence of the approach is demonstrated, and the importance of the background field in the AC losses calculation is underlined. Estimated losses and computing time are compared with the results of the established H-formulation finite element model that includes all the tapes [Zermeno2011]. All calculations were performed using a commercial finite element method (FEM) software [Comsol43a] and a standard desktop computer (Intel i7-3770, 3.40 GHz, 4 cores, RAM 16 GB).

The article is organized as follows. In section II, we define a test case and the reference model that will be used for validation and comparison. In section III, we report our analysis for the homogenized model. In section IV, we perform the same study for the multi-scale model. Finally, estimated losses, computing time and range of application of the models are discussed and compared.

## II. Reference model for validation

### 1) H-formulation

To model superconductors, we use a 2D finite element model with H-formulation [Brambilla2007, Zermeno2011] implemented in COMSOL Multiphysics 4.3a PDE mode application. The superconductor resistivity $\rho$ is modeled by a power law,

$$\rho(\boldsymbol{J}) = \frac{E_c}{J_c(\boldsymbol{B})} \left|\frac{\boldsymbol{J}}{J_c(\boldsymbol{B})}\right|^{n-1} \quad (1)$$

where $J_c$ is the critical current density, $E_c$ is the critical current criteria and $n$ is a material parameter. An anisotropic Kim like model [Kim1962] is used to describe the anisotropic dependence of the critical current density on the magnetic field,

$$J_c(\boldsymbol{B}) = \frac{J_{c0}}{\left(1 + \frac{\sqrt{k^2 B_x^2 + B_y^2}}{B_0}\right)^{\alpha}} \quad (2)$$

where $J_{c0}$, $B_0$, $k$ and $\alpha$ are material parameters. The angle dependence of the critical current density could be easily included. Equation (2) provides a reasonable description of the anisotropic behavior of HTS coated conductors (without artificial pinning) [Grilli2014]. To impose an explicit transport current in each conductor, integral constraints are used. To apply an external magnetic field, appropriate outer boundary conditions are used. Parameters used to simulate YBCO coated-conductors are summarized in Table I. They correspond to a 4 mm tape with a self-field critical current of 99.2 A [Zermeno2013b].

TABLE I
YBCO COATED CONDUCTOR PARAMETERS

| Symbol | Quantity | Value |
|---|---|---|
| $E_c$ | Critical current criterion | $1 \times 10^{-4}$ V/m |
| n | Power law exponent | 38 |
| $J_{c0}$ | Kim model parameter | $2.8 \times 10^{10}$ A/m$^2$ |
| $B_0$ | Kim model parameter | 0.04265 T |
| K | Kim model parameter | 0.29515 |
| $\alpha$ | Kim model parameter | 0.7 |
| $\rho_{air}$ | Air resistivity | 1 Ω.m |

2) **Test case**

To compare the models, we consider a large coil made of 10 pancake coils, each composed of 200 turns (200×10 tapes). We assume here 2D planar geometry, but 2D cylindrical could have been taken. Thanks to symmetries, we can model only one quarter of the coil (100×5 tapes). In the coil cross section, the tapes have an array structure. Let their position be numbered from the bottom left corner by two indices (i,j). The geometric layout and tape numbering are shown in Fig. 2. The layout parameters are summarized in Table II. We study the case of transport current (no applied external field). We impose AC transport currents at 50 Hz with amplitudes of 11 A and 28 A. 11 A is the amplitude at which the central tapes in the outermost column (i=1,j=5) have no virgin regions. This is to say they are filled with either transport or magnetization currents. 28 A is the current at which the central tapes in the outermost column (i=1,j=5) are fully penetrated with transport currents. This is to say that the magnetization currents have been displaced in these tapes. We define this later value as the critical current of the coil since any additional current would imply substantially exceeding the critical current density in at least one tape. We simulate the model for one period and use the second half of said period to compute the average hysteretic losses per cycle (in W/m) using

$$Q = \frac{2}{T} \int_{T/2}^{T} \mathrm{d}t \int_{\Omega} \boldsymbol{E} \cdot \boldsymbol{J} \, \mathrm{d}\Omega \qquad (3)$$

where $T$ is the period of the cycle, and $\Omega$ is the superconductor domain.

TABLE II
TEST CASE LAYOUT PARAMETERS

| Symbol | Quantity | Value |
|---|---|---|
| a | YBCO layer width | 4 mm |
| b | YBCO layer thickness | 1 μm |
| W | Unit cell width | 4.4 mm |
| D | Unit cell thickness | 293 μm |

3) **Reference model**

The reference model is an H-formulation model including all the tapes (Fig.1a). H-formulation models have been widely used and verified against experiments for single conductors and coils [Grilli2007a, Grilli2007b, Nguyen2009, Nguyen2010, Zhang2012, Wang2015, Celebi2015]; and can therefore be trusted to give a good estimation of the real losses. Besides the surrounding air, we model only the superconducting layers taking their real thickness into account. To mesh the superconducting layer, we used a mapped mesh [Zermeno2011] with 100 elements distributed symmetrically following an arithmetic sequence in the width and 1 element in the thickness. We used a mapped mesh between the tapes too. The mesh of one cell of the array is shown in Fig.3.

4) **Results**

The normalized current density distribution and the magnetic flux density distribution at t = 0.015 s (peak current) for the 11 A transport current case are shown in Fig.4; the tape thickness was artificially expanded in the vertical direction for visualization purpose. The magnetic field distribution observed is due to the magnetization currents, which tend to prevent the field to penetrate the tapes. This effect is particularly strong at low transport currents. The hysteretic losses for each tape are plot in Fig.7. In a given row i, the losses spread out across almost 3 orders of magnitude. But in a given column j, the losses are very similar.

### III. Homogenized model

#### 1) Modeling strategy

The homogenized model was introduced in [Zermeno2013b] for modeling 2D stacks of HTS coated conductors (up to 64 tapes). In [Zermeno2014], it was adapted to 3D racetrack coils (50 turns). In comparison to other homogenized models [Clem2007, Yuan2009, Prigozhin2011, Zhang2014], this model does not make any a priori assumptions about the current distribution and allows for overcritical local currents. In this article, we extend it to simulate large arrays of tapes. The homogenized model simulates the coil using equivalent anisotropic homogeneous bulks (Fig.1b) with specifically developed current constraints to account for the fact that each turn carries the same current. This allows us to take advantage of the vanishing edge effect towards the center of the stack (i=1) and to use a very coarse mesh (Fig.6), thus reducing the computing time. For modeling details, we refer the reader to [Zermeno2013b].

#### 2) Results

To provide a qualitative comparison between the reference model and the homogenized model, the 11 A transport current case is analyzed. The normalized current density distribution and the magnetic flux density distribution at t = 0.015 s (peak current) are shown in Fig.5. The distributions calculated with the homogenized model reproduce overall well the reference model. The field distribution in each stack is smoother because the homogenization washes out the actual tapes layout. The hysteretic losses for each tape are plotted in Fig.7. For both transport current cases, the error between the losses estimated with the homogenized model and the reference model is less than 1% (Table III), for a speed up factor of 50-70 (Table IV).

#### 3) Discussion

The step-like shape of the "homogenization" curve is due to the mesh (Fig.6): the mesh is coarse at the center of each stack (i=1) where the spatial variation of the magnetic field is small and finer towards the top of each stack (i=100) to have enough spatial resolution to model the end effects. The coarse mesh helps to reduce the number of degrees of freedom (Table V), and thus the computing time. Locally, the estimated losses follow closely the ones given by the reference model. For the first stack (j=1), the model slightly overestimates them (figure 7). However, this is of little consequence to the overall estimation. We underline the close match for the local losses estimation across the whole coil. This is not trivial as these values span across almost 3 orders of magnitude. The simple implementation of the homogenized model in a commercial finite element program is a clear advantage. Aside from its good ability in predicting hysteretic losses, it proved to be stable. The homogenized model is limited to tape-like conductors with thin superconducting layer. Indeed, in principle the model is equivalent to a stack of infinitely thin tapes and cannot take into account the magnetization currents due to the magnetic field that is parallel to the wide surface of the tape. Furthermore, the homogenization method, at its current stage, cannot include magnetic materials. Thus it is not applicable to Bi-2223 tapes, Bi-2212 or $MgB_2$ wires.

## IV. Multi-scale model

### 1) Modeling strategy

The multi-scale model was introduced in [Queval2013] to estimate the steady-state AC losses of a superconducting wind turbine generator connected to the grid through its AC/DC/AC converter (1058×38 tapes). In this article, we extend it by (a) adding the magnetic field dependence of the critical current density, and (b) introducing a refined background field estimation. This last point permits to include partially the fact that the other tapes are superconducting. The analysis reported here serves as a validation too, as the agreement with the reference model was unclear till now. The main idea behind the multi-scale model is to (1) estimate the background field using a fast coil model, and (2) calculate the AC losses of only several tapes at significant positions of the coil's cross-section using this local field. The losses of the other tapes are then obtained by interpolation. By breaking up the coil domain in several smaller subdomains, we are able to parallelize the problem and thus reduce the computing time. The multi-scale model is composed of two sub-models (Fig.1c). The coil sub-model is a conventional A-formulation magnetostatic finite element model. It includes all the tapes with their actual geometry, and assumes a given current density distribution $J_0$ (uniform or not). The tape sub-model is an H-formulation finite element model. It includes only one conductor with its actual geometry. The coupling between the two sub-models is unidirectional. The magnetic field strength $H$ obtained with the coil sub-model is exported along a coupling boundary lying along the unit cell boundary. It is then applied to the tape sub-model as a time-varying Dirichlet boundary condition. An integral constraint is used to impose the total transport current in the conductor. The meshes used for both multi-scale sub-models are similar to those of the reference model.

### 2) Results

We analyze first the 11 A transport current case. The purpose of the coil sub-model is to obtain a good picture of the local background field to be applied to the tape sub-model.

- *(a) $J_0$ uniform*: In a first approximation, we can consider that $J_0$ is uniform over the tapes cross section (Fig.8a). This is the standard approach [Tonsho2004, Queval2013]. $J_0$ is simply obtained by dividing the coil transport current by the tape cross section area. The resulting magnetic flux density distribution at t = 0.015 s is shown in Fig.8b. The distributions reproduce poorly the ones of the reference model (Fig.4). As a result, the hysteretic losses are underestimated (Fig.10 and Table III). But the speed up factor of up to 700 (Table IV), decreasing the total computing time to less than 90 seconds, might make this model of interest for preliminary design and optimization.

*(b) $J_0$ infinite array*: In reality the local magnetic field of one HTS tape in a coil depends on the adjacent tapes current distribution (both transport and magnetization currents which are not known). In an attempt to include this effect, we propose here to approximate $J_0$ as the current density distribution of an infinite array of 5 tapes (i=∞, j=5) (Fig.9a). $J_0$ is obtained by simulating only 5 tapes side-by-side with appropriate periodic boundary conditions and a coarse mesh. The resulting magnetic flux density distribution at t = 0.015 s is shown in Fig.9b. The distributions now better reproduce the ones of the reference model (Fig.4), especially in the coil center where the infinite array is a good approximation. As expected, the hysteretic losses are in better agreement with the ones obtained from the reference model (Fig.10 and Table III). The error of 6% is considered small bearing in mind that the value of the losses in the tapes spread almost 3 orders of magnitude. However, the preliminary computation of $J_0$ has a cost, both in terms of implementation and computation time (Table IV).

*(c) $J_0$ reference model*: This case has no practical application but demonstrates the convergence of the method. If $J_0$ is taken as the current density distribution of the reference model (Fig.4a), the hysteretic losses are accurately predicted (Fig.10). Therefore the accuracy of the multi-scale model is only limited by the poor estimation of the background field when using $J_0$ uniform or $J_0$ infinite.

In the 28 A transport current case, the multi-scale model with $J_0$ uniform is able to predict the losses with a good accuracy (Table III) for a speed up factor of up to 1600 (Table IV). This can be explained by looking at reference model at t = 0.015 s (Fig.11): the normalized current distribution is almost uniform and produces a field similar to the one of the multi-scale coil sub-model with $J_0$ uniform (Fig.8b).

### 3) Discussion

We calculate only several tapes and interpolate the results to the other tapes. The choice of the tapes is similar to the choice of a mesh. One needs to take more tapes in regions where the field spatial variation is higher. Here it is sufficient to take 5 tapes distributed logarithmically over each stack: the difference with the losses obtained by calculating all the tapes is ~1% (not reported here). The main advantages of the multi-scale model are the low memory requirement (Table V) and the possibility to simulate every tape sub-model in parallel. The computing time in Table IV are given for series computation (1 core), parallel computation (5 cores, e.g. desktop workstation) and full parallel computation (25 cores, i.e. 1 core per tape with dedicated hardware). With $J_0$ uniform, one needs only a static multi-scale coil sub-model. With $J_0$ infinite array, a transient coil sub-model is required, in addition to the preliminary calculation of $J_0$. This explains the difference in computing time in Table IV. For example, for the 11 A transport current case, the computing time for the multi-scale model with $J_0$ uniform was 28 s for the coil sub-model (static), 18 s for export, and 45 s in average per tape sub-model. With $J_0$ infinite array, the computing time was 310 s for the infinite array (transient), 554 s for the coil sub-model (transient), 169 s for export and 87 s in average per tape sub-model. Analytical expressions could be here of interest to estimate $J_0$, and further reduce computing time. We underline that, in comparison to the homogenized model, the multi-scale model could be applied to any kind of conductor, including multifilamentary $MgB_2$, Bi-2223 and Bi-2212 tapes or wires and coated conductor tapes with magnetic substrate. On the one hand, the implementation time of the tape sub-model is low because one needs to model only one tape. However, the coupling between the coil sub-model and the tape sub-model could not be performed directly via COMSOL Multiphysics interface, and required some data processing between exporting and importing the coupling boundary field.

## V.    Conclusion

The motivation behind this work is to develop reliable and fast numerical models to simulate devices wound of hundreds – or even thousands – of HTS coated conductors. Because at least one AC cycle needs to be simulated in order to estimate superconductor AC losses, simulating all the tapes is not an option. We have therefore developed two complementary approaches: the homogenized model and the multi-scale model.

The homogenized model is fast, it provides the best AC losses estimations and it is much easier to set up that the reference model. Being 50-70 times faster than the reference model it provides a good cost-benefit once computing and set up times are considered. Moreover, one could further reduce its computing time at the sacrifice of accuracy by using a coarser mesh. One the other hand, it only works for coated conductor tapes without magnetic substrate. But this is not considered to be a great disadvantage considering that nowadays most coated conductor tapes are manufactured deposited on non-magnetic substrates.

The multi-scale model is the fastest, especially if one takes advantage of parallelization. The use of a coil sub-model with $J_0$ uniform enables very low computing time. But this comes at the price of a larger error, especially for low current amplitude. By improving the quality of the background field (coil sub-model with $J_0$ infinite array), one can reduce the error. But this leads to longer computing time, and relies on the availability of a fast method to approximate the background field for the given problem. In general this model provides a good tradeoff between computing time, number of processors and memory requirements. Besides, it is flexible for design and optimization: once the regions of the coil that contribute the most to the losses are identified, one can study only those regions without the need to compute the whole coil every time. It can in principle be used for any kind of conductor, including multifilamentary tapes or tapes with magnetic materials. It is also easier to set up than the reference model. Finally for extremely large coils, this might be the only method due to the parallel capability.

Both models allow drastic computing time reduction while keeping estimated AC losses in good agreement with those calculated simulating the entire device. Such models could be used to design and optimize large-scale HTS devices.

TABLE III
HYSTERETIC LOSSES

| Model | Losses [W/m] 11 A transport current | Losses [W/m] 28 A transport current |
|---|---|---|
| Reference | 127.02 | 933.99 |
| Homogenized | 127.92 (100.7%) | 934.41 (100.0%) |
| Multi-scale $J_0$ uniform | 96.46 (75.9%) | 909.06 (97.3%) |
| Multi-scale $J_0$ infinite array | 119.85 (94.4%) | 917.57 (98.2%) |

TABLE IV
COMPUTING TIME

| Model | | Time [h] 11 A transport current | Time [h] 28 A transport current |
|---|---|---|---|
| Reference | | 21.3 | 52.0 |
| Homogenized | | 0.3 | 1.1 |
| Multi-scale $J_0$ uniform | Series[a] | 0.3 | 0.7 |
| | Parallel[b] | 0.08 | 0.15 |
| | Full parallel[c] | 0.03 | 0.04 |
| Multi-scale $J_0$ infinite array | Series[a] | 0.9 | 1.4 |
| | Parallel[b] | 0.4 | 0.6 |
| | Full parallel[c] | 0.3 | 0.4 |

a) series computation (1 core), calculated on PC (Intel i7-3770, 3.40 GHz, 4 cores, RAM 16 GB)
b) parallel computation (5 cores), estimated
c) full parallel computation (25 cores), estimated

TABLE V
DEGREES OF FREEDOM

| Model | DOF [] |
|---|---|
| Reference | 606276 |
| Homogenized | 14512 |
| Multi-scale[a] | 1415 |

a) Multi-scale model tape sub-model

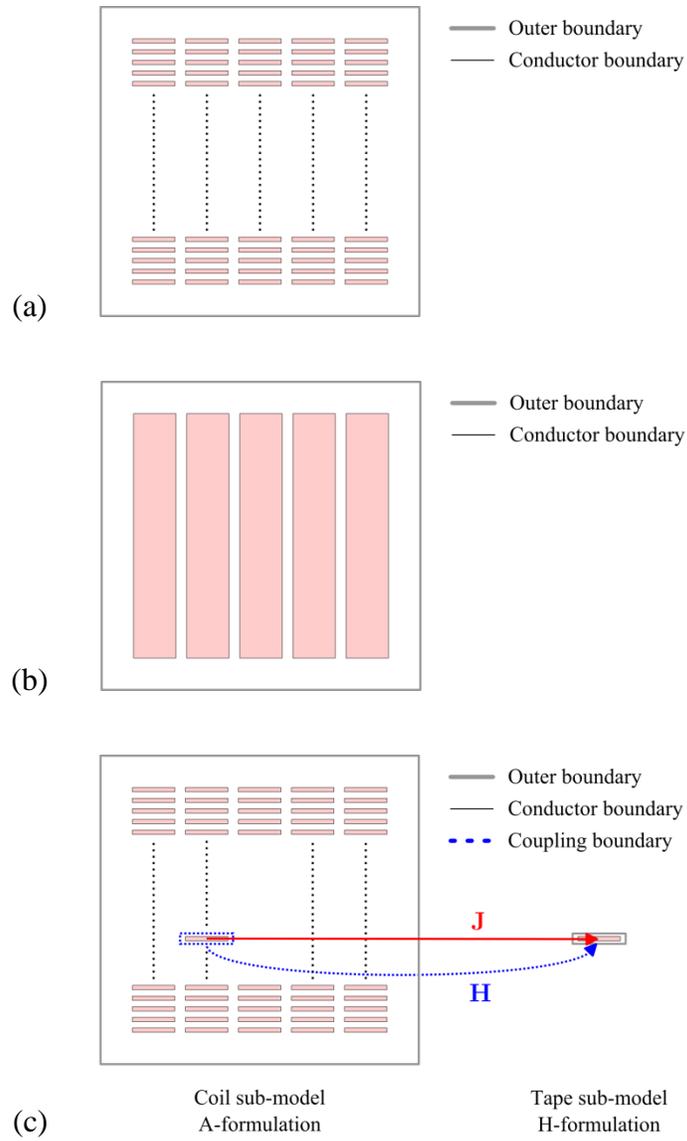

Figure 1 – Overview of the models: (a) Reference model, (b) Homogenized model, (c) Multi-scale model.

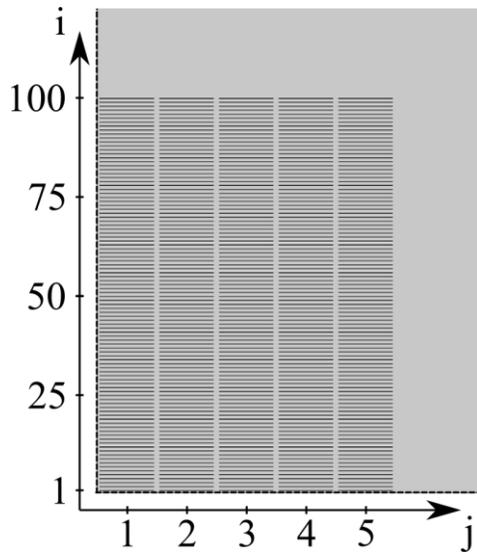

Figure 2 – Test case layout and tape numbering.

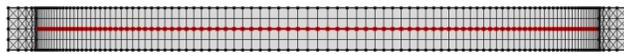

Figure 3 – Unit tape cell mesh.

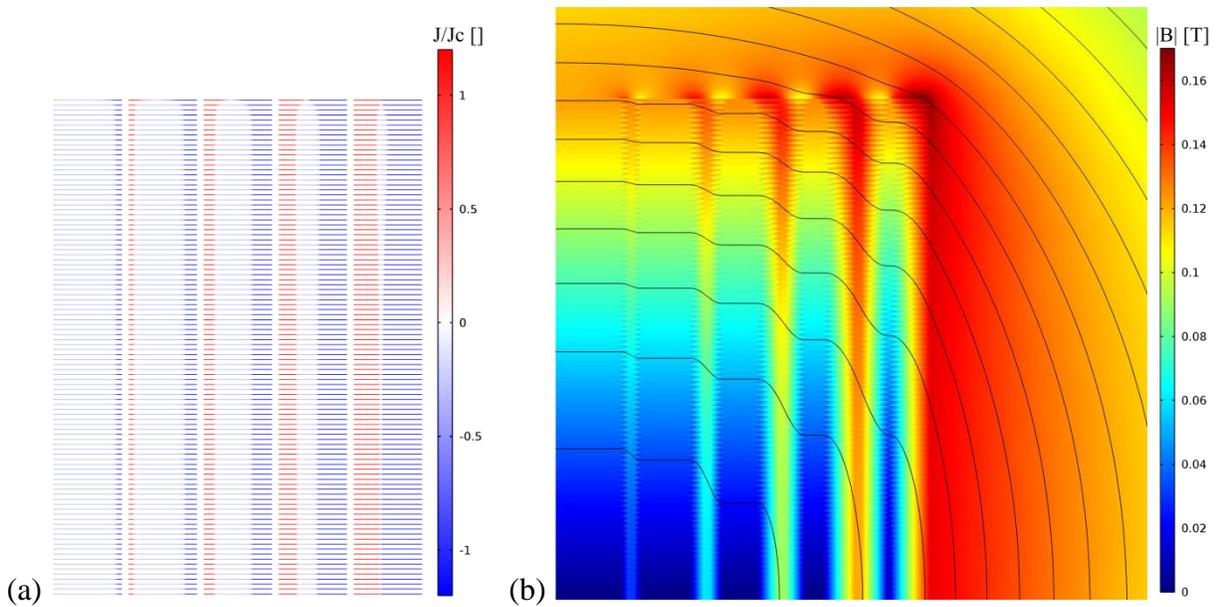

Figure 4 – Reference model (a) normalized current density distribution, (b) magnetic flux density distribution (11 A, t = 0.015 s).

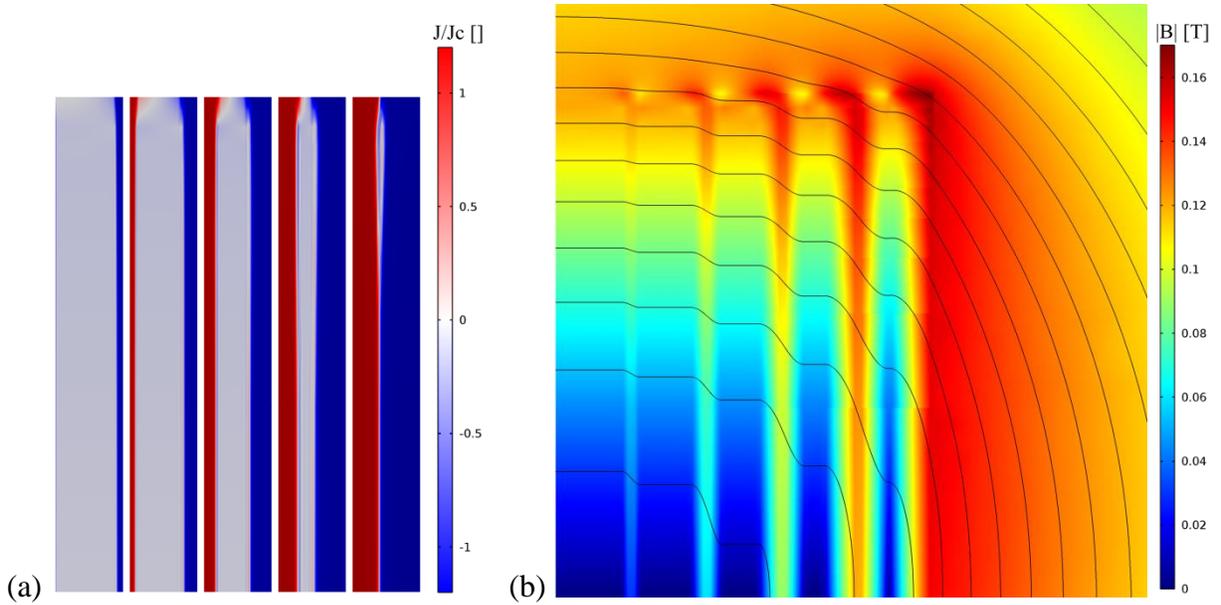

Figure 5 – Homogenized model (a) normalized current density distribution, (b) magnetic flux density distribution (11 A, t = 0.015 s).

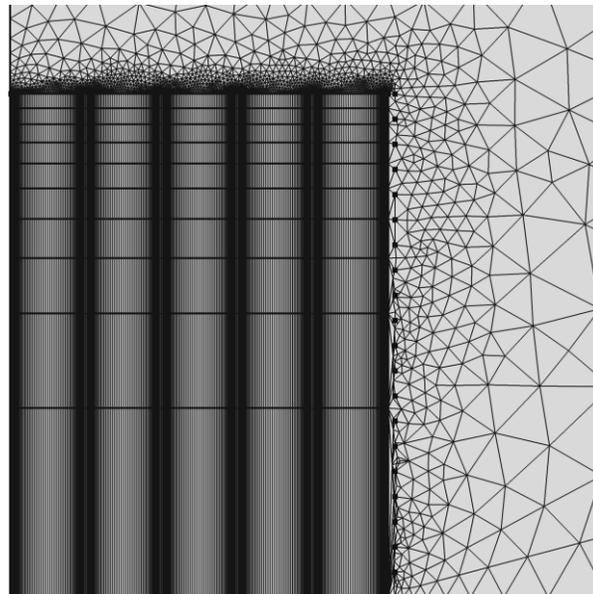

Figure 6 – Homogenized model mesh. Each stack is meshed with 10 elements distributed logarithmically along the height, and 50 elements distributed symmetrically following an arithmetic sequence along the width.

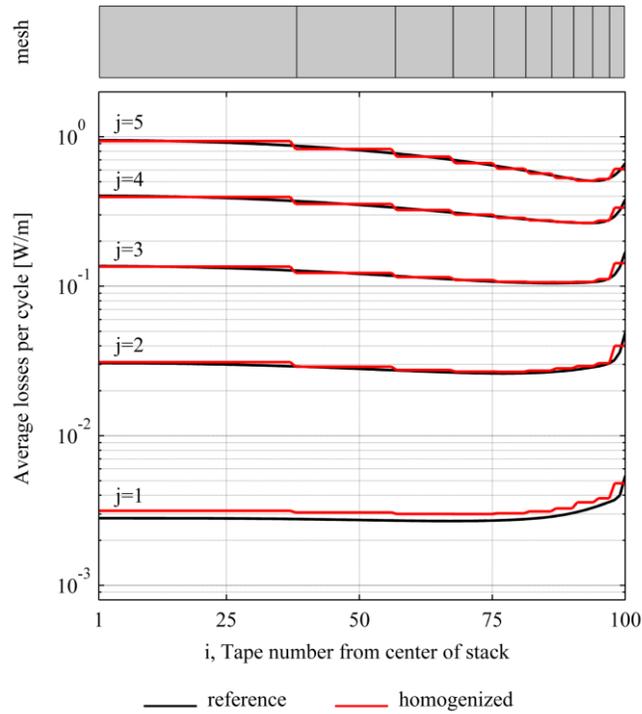

Figure 7 – Homogenized model hysteretic losses for the 11 A transport current case. (i,j) is the tape numbering in the array (see Fig.2). The step-like behavior of the losses is linked to the mesh structure across the height of the stacks as shown above the graph.

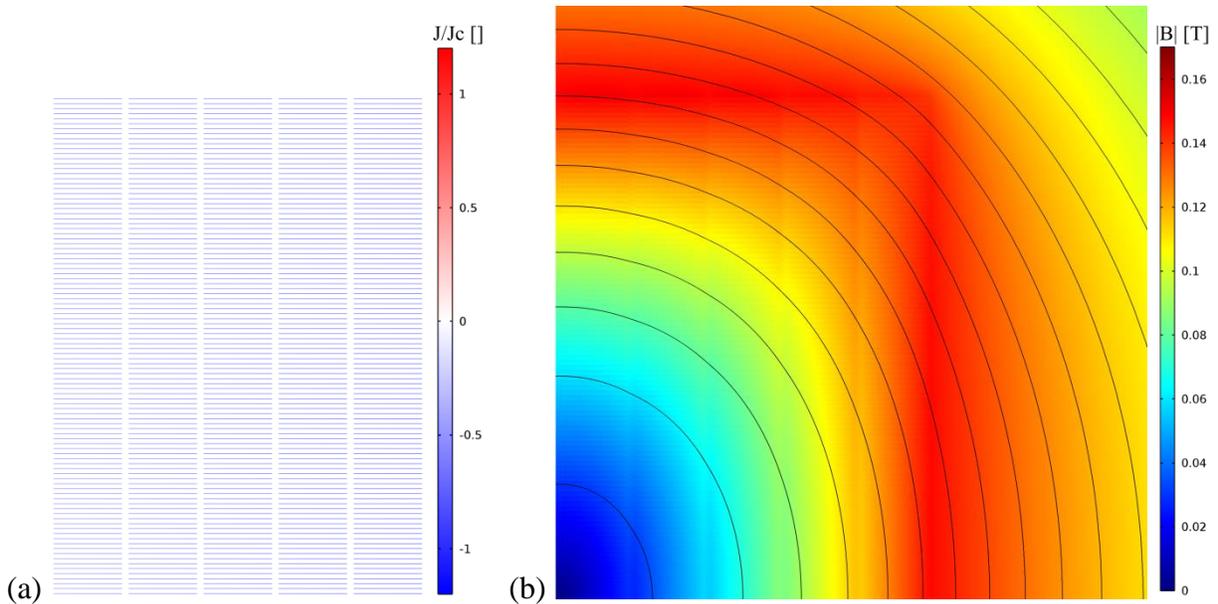

Figure 8 – Multi-scale coil sub-model with $J_0$ uniform (a) normalized current density distribution, (b) magnetic flux density distribution (11 A, t = 0.015 s).

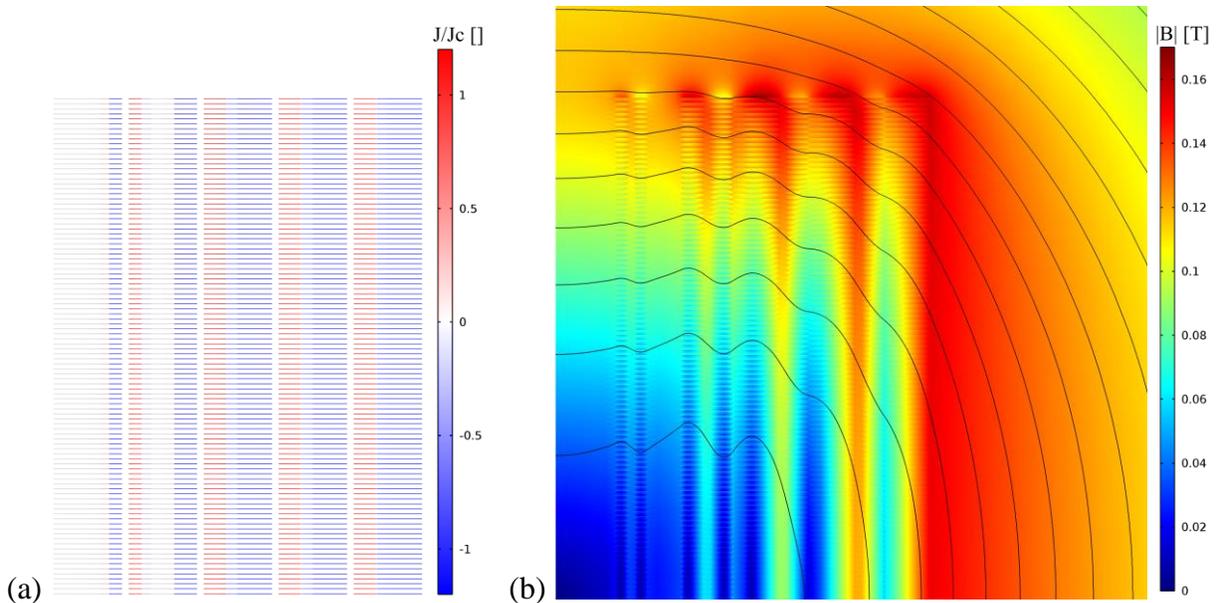

Figure 9 – Multi-scale coil sub-model with $J_0$ "infinite array" (a) normalized current density distribution, (b) magnetic flux density distribution (11 A, t = 0.015 s).

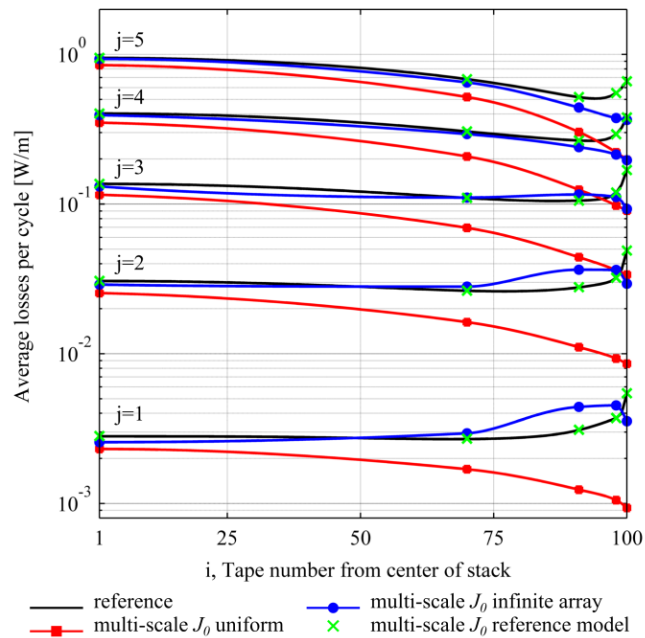

Figure 10 – Multi-scale models hysteretic losses for the 11 A transport current case. The markers show the calculated tapes. Losses for the other tapes are obtained by interpolation. (i,j) is the tape numbering in the array (see Fig.2).

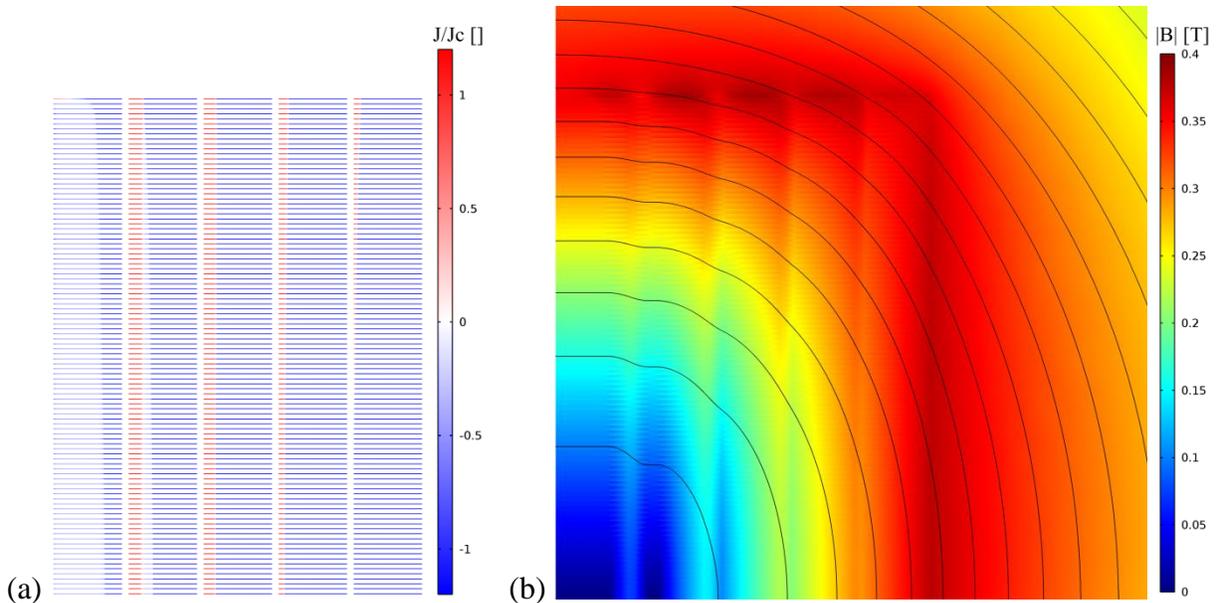

Figure 11 – Reference model (a) normalized current density distribution, (b) magnetic flux density distribution (28 A, t = 0.015 s).